\documentclass[runningheads]{llncs}
\usepackage{graphicx}
\usepackage{comment}
\usepackage{amsmath,amssymb} %
\usepackage{color}
\usepackage{epsfig}
\usepackage{graphicx}
\usepackage{booktabs}
\usepackage[ruled,vlined, linesnumbered]{algorithm2e}
\usepackage{multirow}
\usepackage{verbatim}
\usepackage{dblfloatfix}
\usepackage{wrapfig}
\usepackage{enumitem}
\usepackage{adjustbox}

\begin{document}
\title{Measure Anatomical Thickness from Cardiac MRI with Deep Neural Networks}
\titlerunning{Measure Anatomical Thickness with Deep Neural Networks}
\author{Qiaoying Huang\inst{1}$^{\ddagger}$\thanks{This work was carried out during the internship of the author at United Imaging Intelligence, Cambridge, MA 02140} \and
Eric Z. Chen\inst{2}\and
Hanchao Yu\inst{3}$^{\star}$\and
Yimo Guo\inst{2}\and
Terrence Chen\inst{2}\and
Dimitris Metaxas\inst{1}\and
Shanhui Sun\inst{2}$^{\dagger}$ }
\institute{
Rutgers University, Department of Computer Science, Piscataway, NJ, USA \and
United Imaging Intelligence, Cambridge, MA, USA \and
University of Illinois at Urbana-Champaign, Urbana, IL, USA \\
$^{\ddagger}$ \email{qh55@cs.rutgers.edu}
$^{\dagger}$ \email{shanhui.sun@united-imaging.com}
}

\maketitle              %
\begin{abstract}
Accurate estimation of shape thickness from medical images is crucial in clinical applications.
For example, the thickness of myocardium is one of the key to cardiac disease diagnosis. While mathematical models are available to obtain accurate dense thickness estimation, they suffer from heavy computational overhead due to iterative solvers. To this end, we propose novel methods for dense thickness estimation, including a fast solver that estimates thickness from binary annular shapes and an end-to-end network that estimates thickness directly from raw cardiac images.We test the proposed models on three cardiac datasets and one synthetic dataset, achieving impressive results and generalizability on all. Thickness estimation is performed without iterative solvers or manual correction, which is $100\times$ faster than the mathematical model. We also analyze thickness patterns on different cardiac pathologies with a standard clinical model and the results demonstrate the potential clinical value of our method for thickness based cardiac disease diagnosis.

\end{abstract}

\section{Introduction}
Estimation of shape thickness from images is a fundamental task and has wide applications in practice.
Particularly, in the medical domain, anatomical thickness plays a crucial role in disease diagnosis.
For instance, abnormal changes of left ventricular myocardial (muscle wall) thickness are signs of many heart diseases~\cite{khalifa2011accurate,sliman2014assessment,yu2020foal,yang2020mri}.

This work specifically focuses on predicting the thickness of left ventricular walls, which can be acquired from cardiac Magnetic Resonance Imaging (MRI)~\cite{yang2018multi,yu2020motion,Wu2020cardiac}.
There are two common ways to measure such heart anatomical thickness.
The first method is to measure regional wall thickness (RWT)~\cite{xue2017direct,xue2018full}, which divides the heart into several regions and respectively averages the thickness to minimize the effort of manual measurement on the entire heart.
RWT only provides average values of myocardial regions, and hence has limitations in helping physicians understand precise thickness change for clinical diagnosis.
A better solution is dense wall thickness (DWT)~\cite{jones2000three,yezzi2003eulerian,khalifa2011accurate,sliman2014assessment}, which provides detailed thickness measurements at every location of the entire heart. 
It is also capable of dealing with some convoluted surfaces, which is not solvable by simply calculating the Euclidean distance between corresponding points of the inner and outer borders~\cite{jones2000three}.
However, this approach suffers from a heavy computational burden since numerical methods are used to solve a second-order partial differential equation (PDE).
For example, it takes approximately one minute to estimate the thickness of the brain cortex in a 3D volume \cite{yezzi2003eulerian}.
Moreover, inner and outer contours need to be defined manually in some disconnected cases, which is time-consuming. 
Therefore, it is not feasible to apply this method for variously folded and faulted structures. An efficient and effective approach is highly desired.

\begin{figure}[!t]
    \centering
    \includegraphics[width=0.7\linewidth]{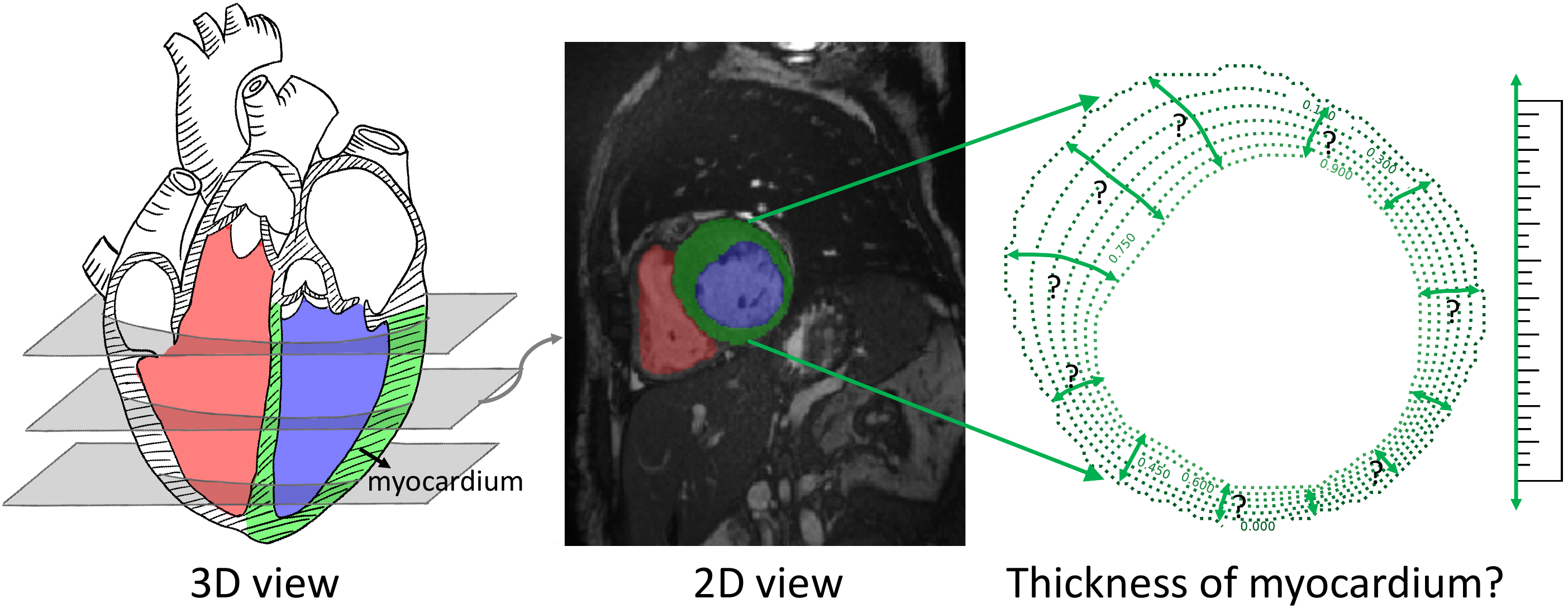}
    \caption{Overview of 2D multi-slice cardiac MR data acquisition (left) and myocardium thickness measurement (right) from a corresponding 2D MR image (middle). }
    \label{fig:my_label}
\end{figure}

To the end, we consider estimating DWT directly from raw images in an automatic manner.
However, it is non-trivial to derive an effective approach without utilizing shape information since the model needs to implicitly learn the shape to predict the thickness of the heart.  
Additionally, segmentation and thickness estimation are highly entangled.  
Once the segmentation changes, the thickness estimation has to adapt to a new shape from the last segmentation prediction, making the training hard to reach an equilibrium.
Therefore, we start with a simpler problem that estimates thickness from a binary image (shape). 
Inspired by the recently proposed deep PDE solver~\cite{hsieh2019learning}, we introduce a deep learning-based thickness solver.
The thickness estimation problem is more challenging than~\cite{hsieh2019learning}, since additional procedures are involved, such as integrating lines and interpolating missing values (refer to section~\ref{sec:math}).
Base on this solver, we then propose a novel end-to-end network for the original problem, which effectively decomposes the complex process into relatively easy sub-processes and utilizes shape information learned from the segmentation to benefit the thickness estimation.

Our major contributions are four aspects.
(1) We introduce a fast thickness solver that estimates thickness from binary images and train it with a synthetic dataset, making it more generalizable to unseen shapes.
(2) We propose a novel end-to-end network that predicts thickness directly from raw images.
(3) We conduct a comprehensive thickness pathology analysis to show the potential clinical values of the proposed framework.
(4) To the best of our knowledge, this is the first DWT deep learning-based approach without hand-engineered steps nor manual annotations.

\section{Method}
\subsection{Background}\label{sec:math} 
\begin{figure*}[!t]
    \centering
    \includegraphics[width=0.8\textwidth]{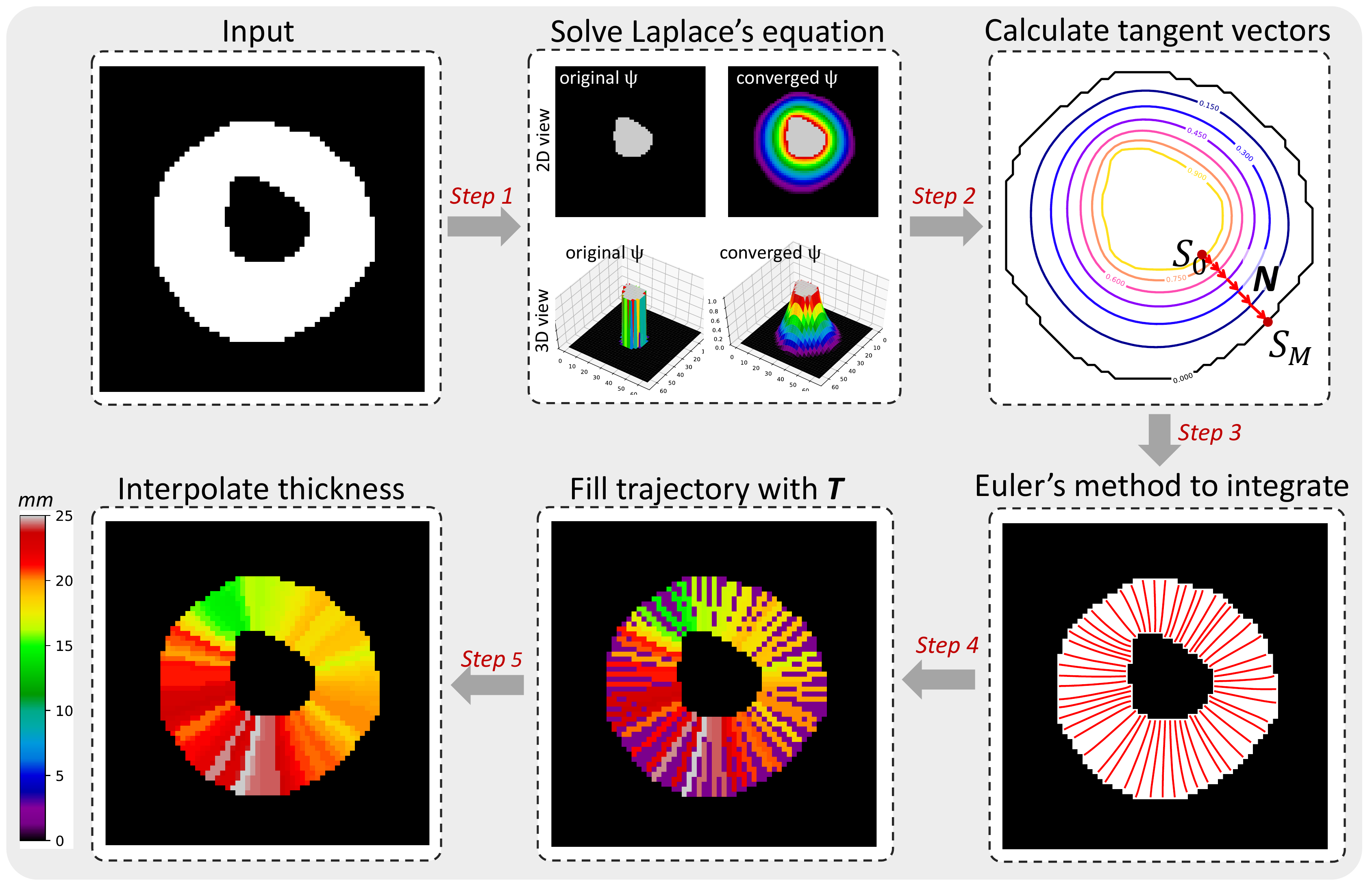}
    \caption{Overview of dense thickness computation using the mathematical model. 
    Given any shape, e.g., an annular binary mask, we first solve Laplace's equation to obtain the equipotential surfaces (contour lines in step 2). 
    Then we calculate tangent vectors (red arrows in step 2) that are orthogonal to each surface. 
    Euler's method is used to integrate the tangent vectors (red lines in step 3). 
    Finally, an interpolation algorithm is performed to estimate the missing values (purple pixels in step 4) based on the potential surface and adjacent known thickness values.}
    \label{fig:pipeline}
\end{figure*}
We first introduce the mathematical model for thickness estimation based on~\cite{jones2000three}, which is used to generate training data for the proposed deep learning models. 
Suppose the input is a binary shape as in Fig.~\ref{fig:pipeline}.  
We consider the electric potential value at the inner contour $S$ to be the maximum ($\psi$) and that at the outer contour $S'$ to be the minimum ($0$), where $\psi>0$.
Laplace's equation is a second-order PDE that defines the electric potential function over a region enclosed by boundaries $S$ and $S'$. 
Mathematically, for a 2D surface, Laplace's equation is: 
\begin{equation}\label{eq:laplace}
    \triangledown ^{2}\psi = \frac{\partial^{2}{\psi}}{\partial x^{2}} + \frac{\partial^{2}{\psi}}{\partial y^{2}} = 0,
\end{equation}
which is usually solved by numerical methods~\cite{numerical1992press}.
Step 1 of Fig.~\ref{fig:pipeline} shows the 2D and 3D views of the initial $\psi$ and the converged $\psi$.

The next step is to compute the vector $N=(N_x, N_y)$ orthogonal to each potential surface, as shown in step 2 of Fig.~\ref{fig:pipeline}.
The path length of these integrated tangent vectors is the thickness between two corresponding points.
Suppose $S_0=(S_0(x),S_0(y))$ is the starting point at the inner border.
A curve $S_0S_1S_2\cdots$ is formed by iteratively taking small steps $d_s$ along the tangent line until point $S_n$ touches the outer border.
We use Euler's method~\cite{an1989kendall} to integrate:
\begin{equation}\label{eq:euler}
S_{n+1}(x) = S_n(x) + N_x \cdot d_s, \quad S_{n+1}(y) = S_n(y) + N_y \cdot d_s. 
\end{equation}
All integrated lines are plotted in step 3 of Fig.~\ref{fig:pipeline}.
The thickness is computed as
$T = \sum_{n=1}^{M}\sqrt{(S_n(x)-S_{n-1}(x))^2+(S_n(y)-S_{n-1}(y))^2}$ for $M$ points along the trajectory.
Since the number of points at the inner boundary must be less than that of the outer boundary, the points with unknown thickness values need to be filled (the dark purple points in step 4 of Fig.~\ref{fig:pipeline}). 
We propose to interpolate these points by using the known thickness points and the potential value $\psi$.
Step 5 of Fig.~\ref{fig:pipeline} presents the final thickness result after interpolation.
The speed of the mathematical model depends on thickness of a structure, the number of points at the inner border and the step size $d_s$ in Eq.~\eqref{eq:euler}.
It is often slow and needs manual correction.
To address the above drawbacks, we propose deep learning-based approaches for thickness estimation.

\subsection{Thickness Computation via a Fast Solver}\label{sec:solver}
We first focus on the thickness estimation that takes a binary image as input.
We propose a fast solver to replace the mathematical model.
The solver adopts a U-Net architecture~\cite{ronneberger2015u}, denoted as $G$.
It takes a \textit{binary image} $s$ as input and estimates the thickness $\hat{y}$, represented as $\hat{y}=G(S)$.
The goal is to minimize the difference between predicted thickness and ground truth thickness $y$, which is calculated from the mathematical model.
This can be achieved by optimizing over an $l_2$ loss imposed on $\hat{y}$ and $y$ along with an $l_1$ regularization on $\hat{y}$:
\begin{equation}\label{eq:ltloss}
\ell_\mathrm{thick}(y, \hat{y})=\|y-\hat{y}\|_{2}^{2}+\alpha\|\hat{y}\|_{1},
\end{equation}
where the $l_1$ is used to force sparsity on $\hat{y}$, since the thickness map exhibits many zeros in background.
$\alpha$ is a weight to balance two losses.
With a large number of diverse training data, the solver generalizes to even unseen shape and achieves $100\times$ speedup compared to the mathematical model.

\begin{figure}[!t]
    \centering
    \includegraphics[width=0.7\textwidth]{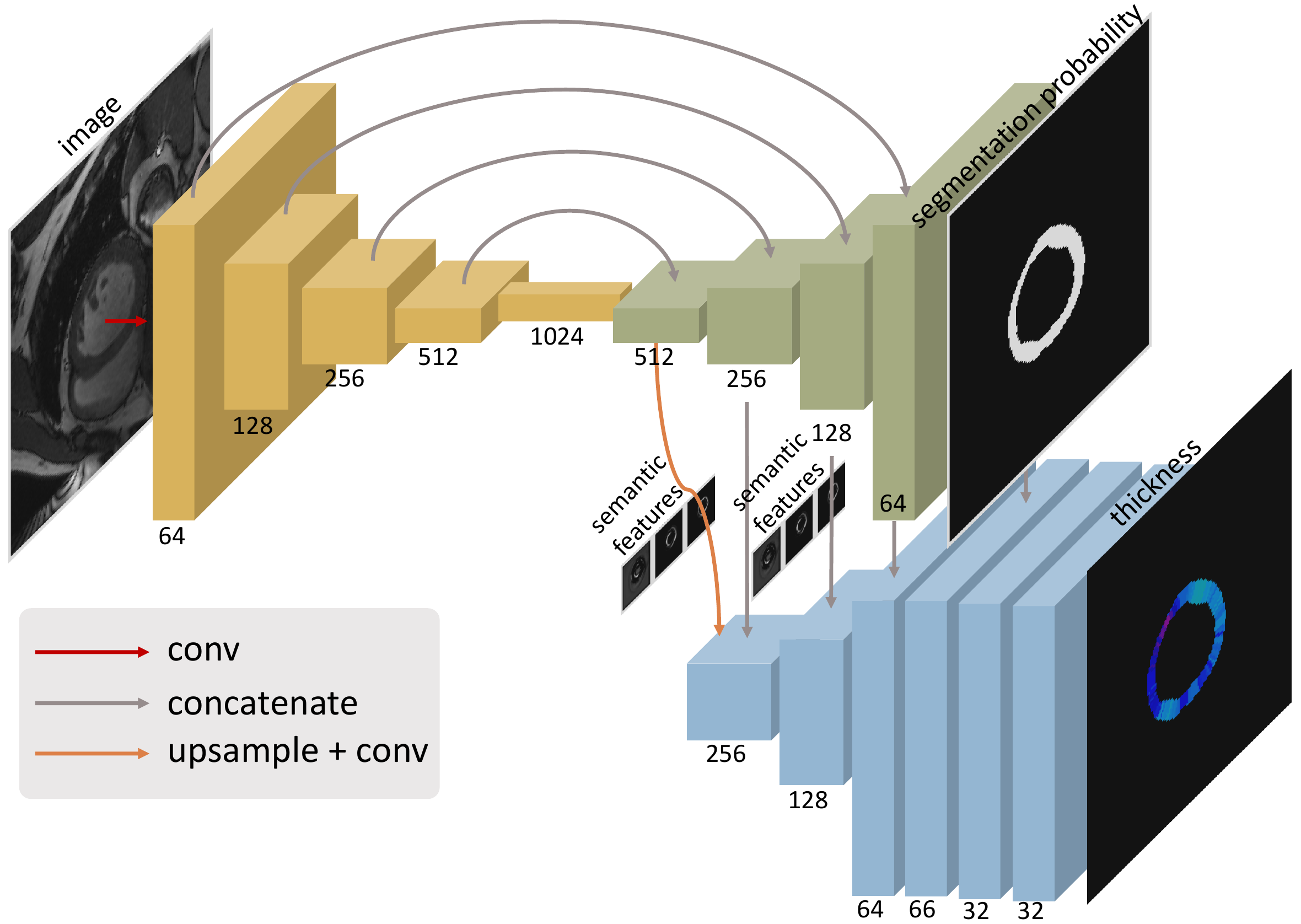}
    \caption{Illustration of the proposed end-to-end network for thickness estimation directly from the raw image.}
    \label{fig:network}
\end{figure}

\subsection{Thickness Computation via an End-to-end Network}\label{sec:solver_image}
In this section, we propose an end-to-end network that predicts thickness directly from the \textit{raw image}, as illustrated in Fig.~\ref{fig:network}.
The network perceives the essential shape information during thickness estimation by learning the segmentation mask of the heart.
It effectively disentangles shape understanding (segmentation) and thickness estimation tasks.
The problem now becomes taking a raw image $x$ as input and predicting both segmentation (shape) $s$ and thickness $y$.
To leverage semantic information from the segmentation $s$, we enforce the decoder of the thickness estimator to utilize features generated from the segmentation decoder via feature map concatenation. 
The thickness $y$ is generated based on $s$ that learns features directly from $x$. 
The reason why decomposing the thickness estimation task into relatively easy sub-processes is that it is non-trivial to directly predict dense thickness from a raw image since it needs to infer a shape implicitly, which has proven by section~\ref{sec:math} and~\ref{sec:solver}. 
In section~\ref{sec:exp}, we will show the results of training a U-Net with the loss function in Eq.~\eqref{eq:ltloss} are blurry with notable artifacts.
The U-Net attempts to regress a wide range of float point numbers without using shape information but obtains the mean of them. 
Therefore, to address the above limitations, we propose our novel end-to-end network that utilizes the semantic features from the segmentation task.

\begin{figure}[!t]
    \centering
    \includegraphics[width=0.6\linewidth]{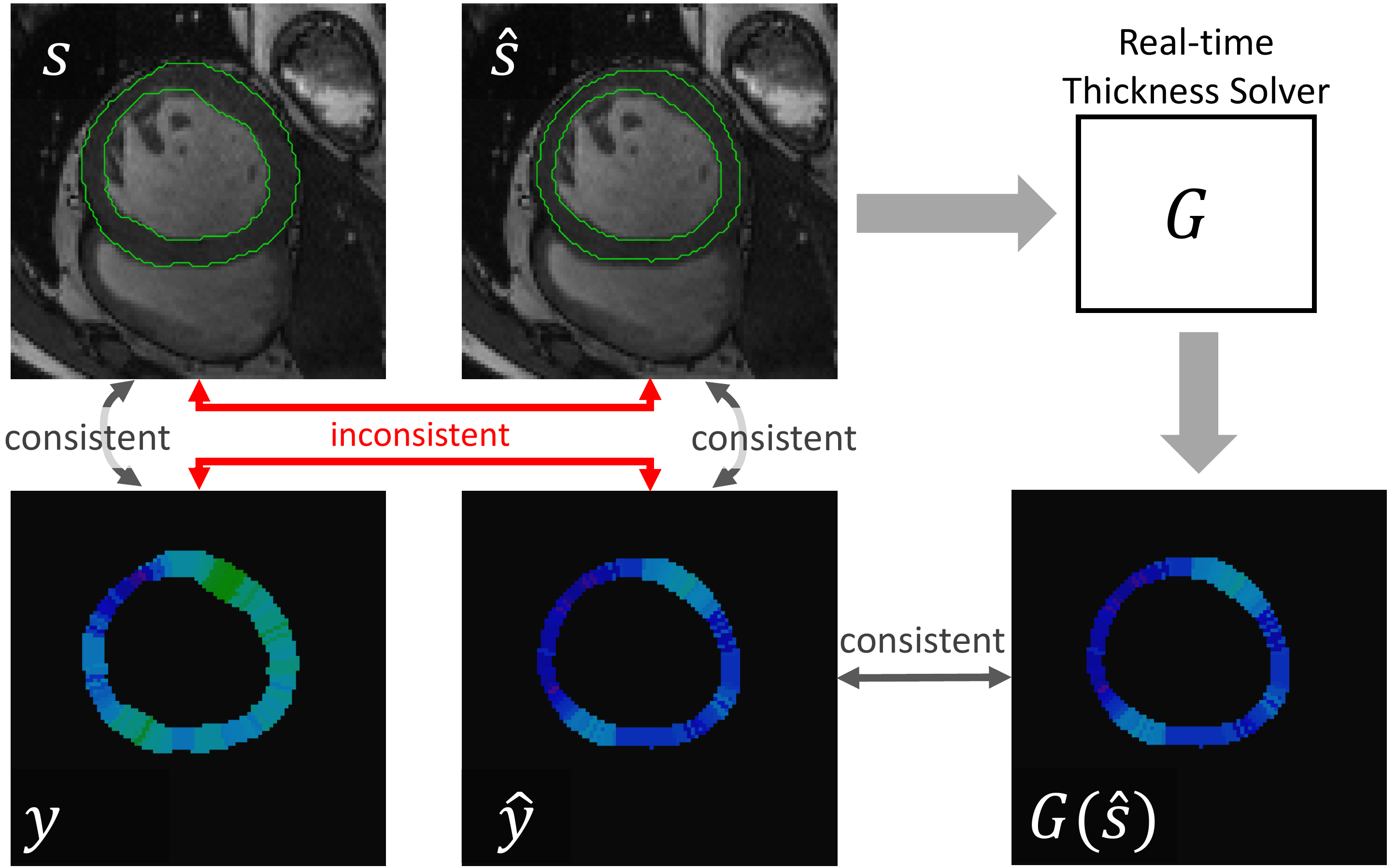}
    \caption{Illustration of the shape inconsistency problem between predicted thickness $\hat{y}$ and ground truth $y$.
    Note that the underline predicted shapes $\hat{s}$ and ground truth shape $s$ are different. 
    During training, we replace the ground truth $y$ using the proposed generic fast thickness solver $G$ to remove this shape discrepancy.}
    \label{fig:unpair}
\end{figure}

Another challenge is to find an optimal loss function. 
By training the model with a combined segmentation loss and thickness loss, we found that segmentation dominates the training and thickness estimation is not able to converge. This is caused by the shape inconsistency problem (Fig.~\ref{fig:unpair}), where the predicted shape $\hat{s}$ is inconsistent with the shape of the thickness ground truth $y$.
Since the thickness decoder utilizes shape features from $\hat{s}$ to predict thickness $y$, it struggles between two different shapes between $\hat{s}$ and $y$, making it difficult to reach an equilibrium of training. 
Therefore we can not use the ground truth thickness $y$ as the guidance for predicting thickness.
The solution is to replace $y$ with $G(\hat{s})$ as the ground truth for thickness estimation since $G(\hat{s})$ has the same shape as $\hat{s}$, where $G(\hat{s})$ is the generated thickness from the estimated mask $\hat{s}$ via the proposed fast thickness solver $G$. 
We aim to minimize the following loss: 
\begin{equation}\label{eq:newloss}
    \mathcal{L}=\ell_\mathrm{thick}(G(\hat{s}), \hat{y})+\beta \ell_\mathrm{seg}(s, \hat{s}),
\end{equation}
where 
$\ell_\mathrm{seg}$ is cross entropy loss for segmentation and $\beta$ is to balance two terms.

\begin{figure}[!t]
    \centering
    \includegraphics[width=0.75\textwidth]{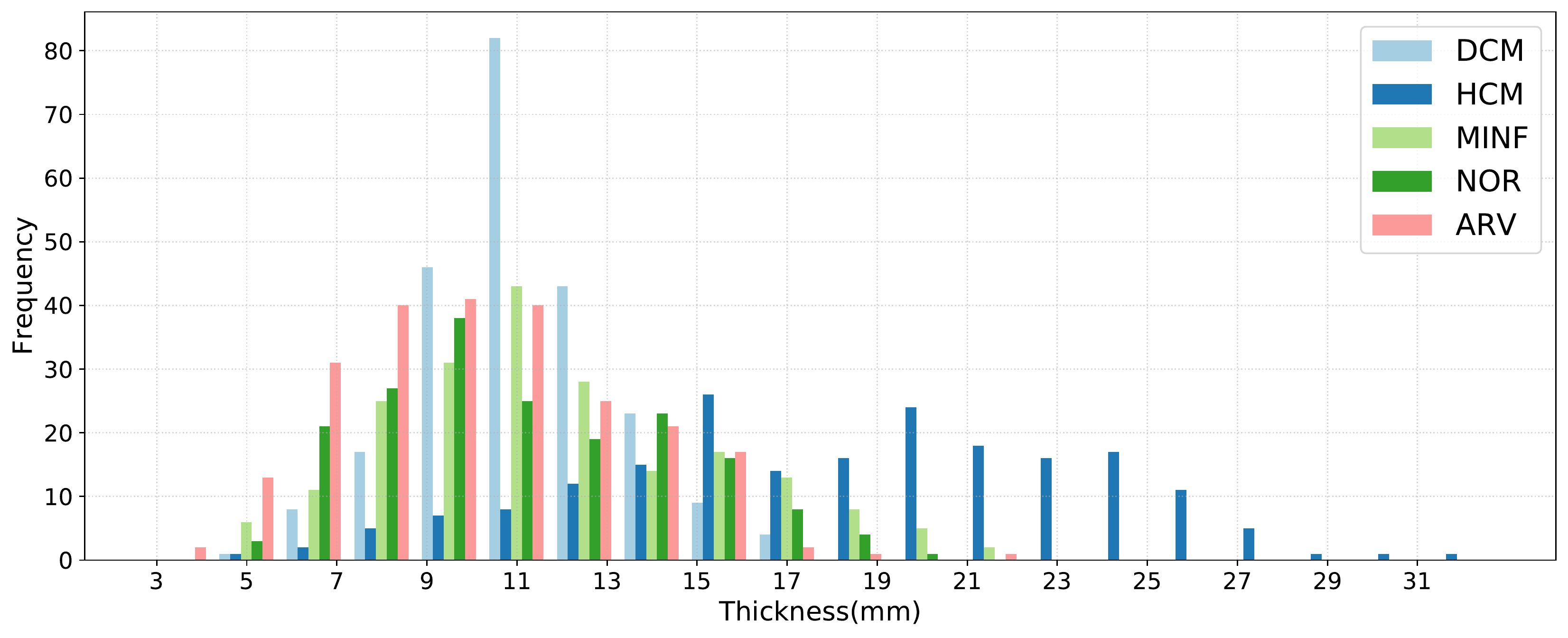}\\
    (a)\\
    \includegraphics[width=0.75\textwidth]{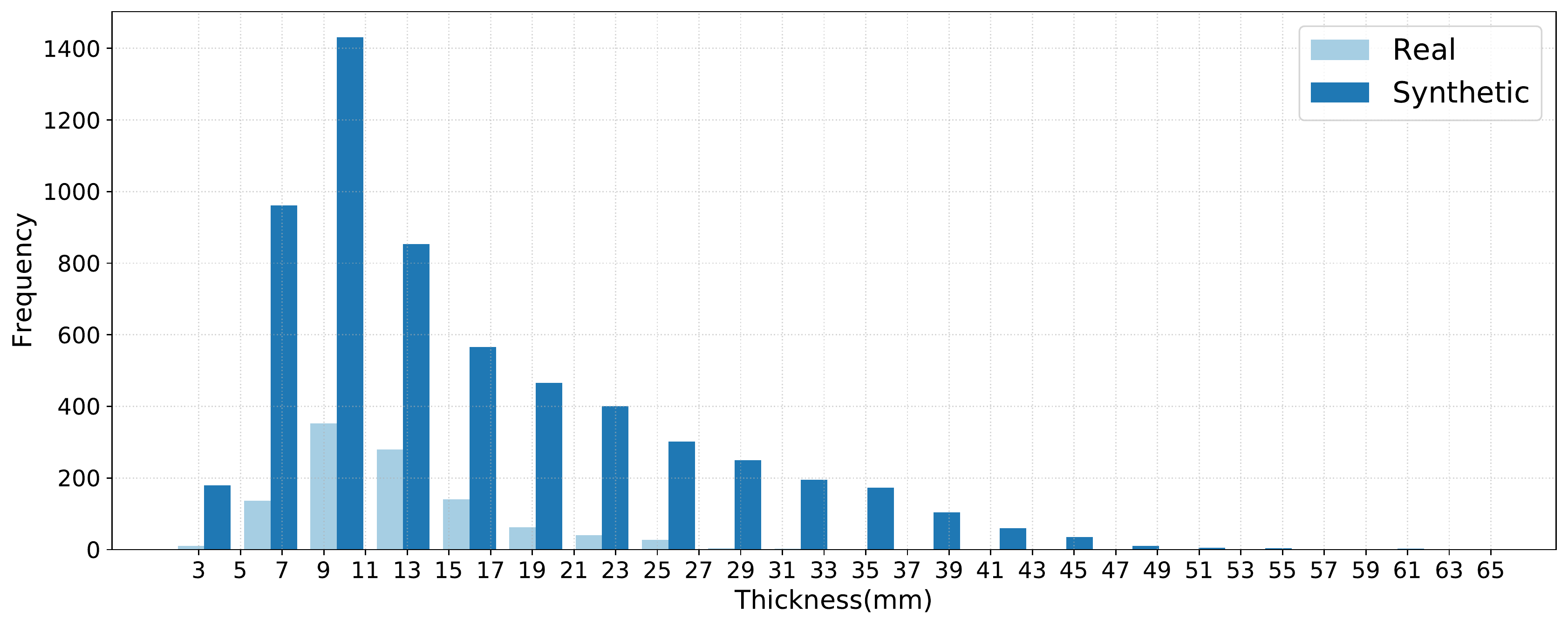}\\
    (b)
    \caption{(a) Maximum thickness value distribution of five cardiac diseases from ACDC data.(b) Maximum thickness value distribution from our synthetic data. For a comparison, the corresponding distribution from ACDC data (Real) is also included.}
    \label{fig:distrib}
\end{figure}

\section{Experimental Results}\label{sec:exp}
For model training, we mainly experiment on two following datasets: 
(1) \textit{ACDC} dataset \cite{bernard2018deep} demonstrates various thickness patterns of different heart diseases, which contains 100 subjects (1,742 2D slices after removing the outliers) and includes five pathologies: dilated cardiomyopathy (DCM), hypertrophic cardiomyopathy (HCM), myocardial infarction (MINF), abnormal right ventricle (ARV) and normal people (NOR).
We plot the maximum thickness value distribution of the five categories in Fig.~\ref{fig:distrib}(a).
We observed that the maximum thickness of each 2D image ranges from 3 to 32 $mm$.
DCM has the largest number of thickness values around 9-11 $mm$.
This is consistent with the fact that DCM is a disease with an enlarged and less functional left ventricular than normal people.
HCM patient has thicker myocardium in diastole in several myocardial segments compared to a healthy person.
We randomly separated the 100 patients into training (60 patients, 1,016 slices), validation (20 patients, 388 slices) and test (20 patients, 338 slices) set with equal ratio of five pathologies in each dataset.
The pixel spacing is normalized to 1.36 $mm$ and the image size is adjusted to 
$192\times192$ by center cropping and zero padding.
We applied the mathematical model in Section~\ref{sec:math} to ACDC data and 
analyze the thickness pattern in each disease category. 
(2) \textit{Synthetic} dataset is built to cover more general and diverse binary annular shapes, as Fig~\ref{fig:distrib} (b) shown. 
The synthetic dataset contains about 6,000 images and ranges from 0 to 64 $mm$ that has a wider distribution compared to ACDC.
We propose the synthetic dataset since the input of the fast thickness solver are binary images, it is easy to synthesize.
We also want to show the solver has a good generalizability and can be trained without the burden of annotating data. 
The synthesis process is to first generate binary annular shape with random radius and position, and then apply elastic transform and piecewise affine transform with random parameter settings.
After generating the synthetic dataset, we also applied the mathematical model to generate the ground truth thickness value. 

All models are trained with Adam optimizer with learning rate of $\mathrm{10}^{-4}$ and $350$ epochs.
We set $\alpha=\mathrm{10}^{-3}$ in Eq.~\eqref{eq:ltloss}.
The evaluation metrics are Mean Absolute Error (MAE) and Mean Square Error (MSE) inside myocardium regions. 

\begin{table}[!b]
    \centering
     \renewcommand{\tabcolsep}{4pt}
    \caption{Results of the proposed thickness solvers on ACDC and synthetic datasets.
    }
    \begin{tabular}{llcclcc}
    \toprule
      \multirow{2}{*}{Model}   &&  \multicolumn{2}{c}{ACDC} && \multicolumn{2}{c}{Synthetic} \\
      \cmidrule{3-4} \cmidrule{6-7}
      && MAE ($mm$) & MSE ($mm$) && MAE ($mm$) & MSE ($mm$)\\
      \midrule
      Au-En && 1.061(0.284) & 3.472(2.665) && 0.815(0.156) & 2.219(1.253) \\
      U-Net-$2$ && 0.349(0.067) & 0.214(0.096) && 0.335(0.061) & 0.198(0.090) \\
      U-Net-$3$ && 0.355(0.072) & 0.221(0.109) && 0.322(0.058) & 0.185(0.082) \\
      U-Net-$4$ && 0.350(0.069) & 0.212(0.104) && \textbf{0.321(0.060)} & \textbf{0.185(0.085)} \\
     \bottomrule
    \end{tabular}
    \label{tab:realsyn}
\end{table}

\begin{figure}[!t]
    \centering
    \includegraphics[width=0.85\textwidth]{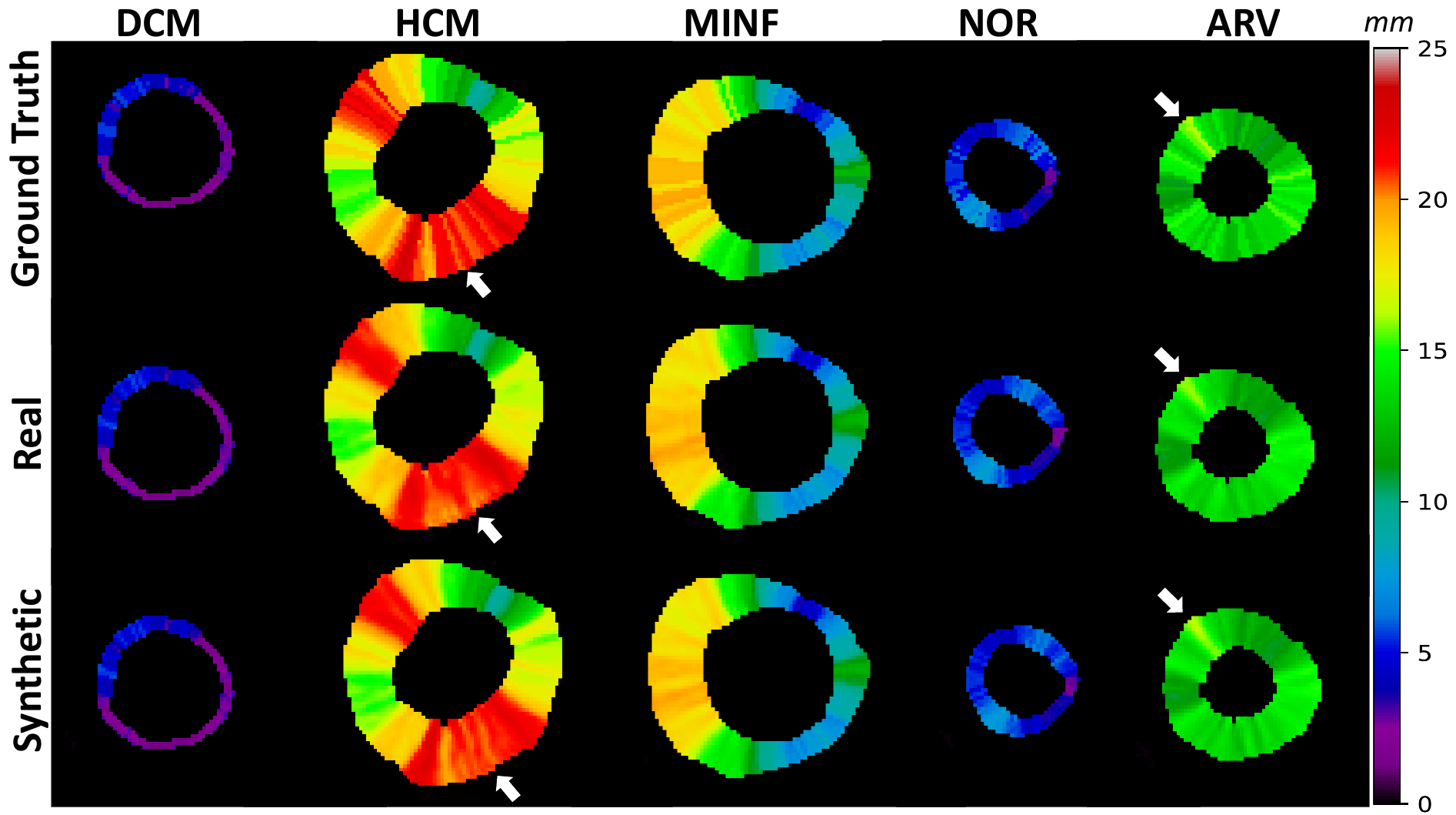}
    \caption{
    Examples of thickness prediction by U-Net-$4$ that are trained on real ACDC data or synthetic data.
    Tests were performed on ACDC test data including four disease categories (DCM, HCM, MINF, ARV) and one normal category (NOR). 
    The white arrows indicate the area that the U-Net-$4$ trained with synthetic data performs better than with Real data. 
    }
    \label{fig:rts}
\end{figure}

\begin{figure}[!t]
    \centering
    \includegraphics[width=\textwidth]{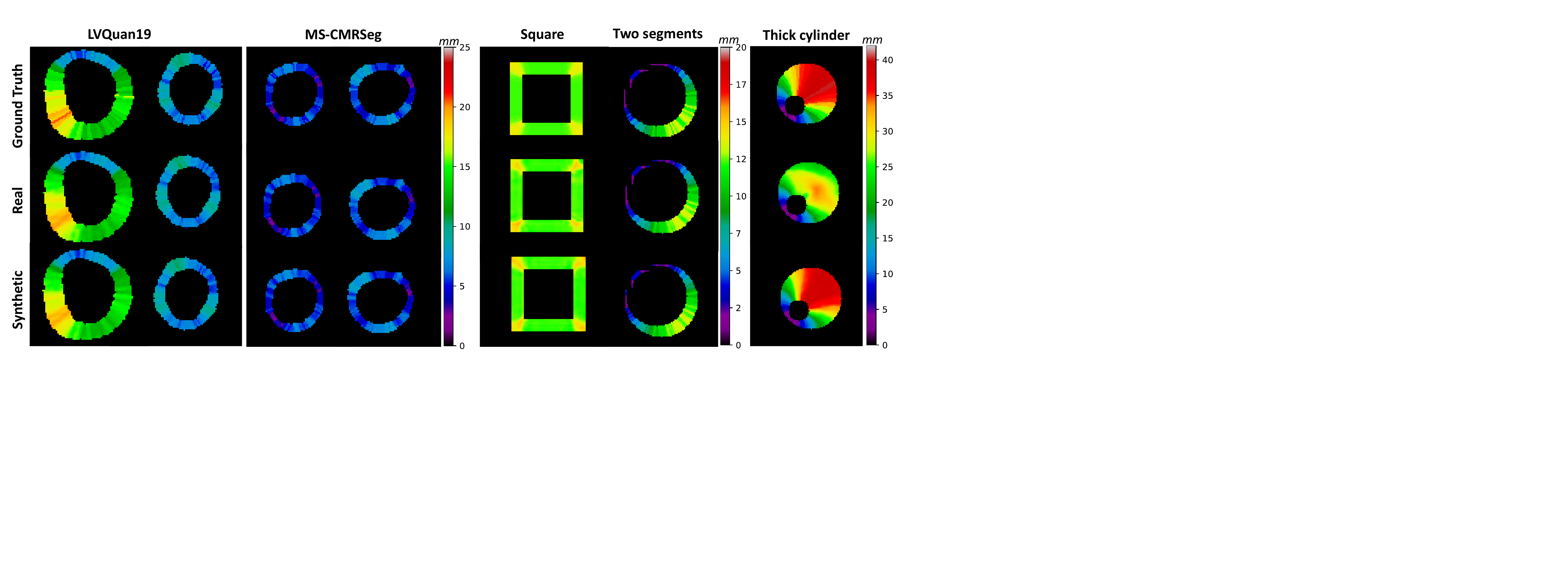}
    \caption{Examples of thickness prediction on unseen data with the fast thickness solver.
    }
    \label{fig:gen}
\end{figure}

\paragraph{\textbf{Thickness Estimation on Binary Image}}\label{sec:realsolver}

Since there is \textbf{NO} other deep learning-based dense thickness estimation method, we consider the Auto-Encoder as a baseline model. 
U-Net-$k$ as our thickness estimation models ($k$ is the number of downsampling and upsampling blocks).
These models are trained on real data and synthetic data, respectively. 
For a fair comparison, all models are evaluated on the same ACDC test set.

The results of thickness estimation are reported in Table~\ref{tab:realsyn}.
U-Net-$k$ models outperform Auto-Encoder in MAE and MSE, demonstrating that the skip-connection is essential in thickness estimation.
We did not report models with deeper layers since no further improvement compared to the U-Net-$4$. 
Furthermore, models trained on synthetic data work better than those with real data, indicating the benefit of including more diverse cases.
Fig.~\ref{fig:rts} presents examples of thickness estimation with different cardiac pathologies from models trained on real data and synthetic data, respectively.
The model trained on synthetic data produces more similar results to the ground truth, especially in the cases of DCM and MINF. 
Even in the most challenging case of HCM, the model with synthetic data training recovers better details.

To validate the generalizability of our proposed thickness model, we extended evaluations on two unseen real test datasets and one unseen synthetic test dataset. 
The real datasets are \textit{LVQuan19}~\cite{LVquan19} and \textit{MS-CMRSeg}~\cite{zhuang2018multivariate}, where manual annotations of the myocardium are provided.
There are 56 normal subjects with $1,120$ 2D images in \textit{LVQuan19} and $45$ normal subjects with $270$ 2D images in \textit{MS-CMRSeg}.
The synthetic test dataset (\textit{Gen-Data}) includes three additional shapes that are unseen in the training set: square and thick cylinder and two segments.
The validation results are reported in Table~\ref{tab:gen}.
Both models trained on real data (ACDC) and synthetic data respectively achieve satisfying accuracy, demonstrating that our proposed thickness solver has good generalizability. 
The model trained on synthetic data outperforms the one on real data in terms of MAE and MSE. 
Fig.~\ref{fig:gen} illustrates the validation results. 
The solver trained on the synthetic data performs well on all tested data sets while the solver trained on the ACDC data fails on extremely thick structures (16.0~$mm$ MSE in Table~\ref{tab:gen}, refer to the thick cylinder example in Fig.~\ref{fig:gen}). 
We conclude that the solver trained on a large amount of synthetic data is robust and can be utilized as a generic thickness solver. 
This mitigates the data scarcity problem for training a deep learning based thickness solver.

\begin{table}[!b]
\begin{minipage}{0.49\linewidth}
    \centering
    \setlength{\tabcolsep}{5pt}
    \caption{Results of model generalizability.
    }
    \label{tab:gen}
    \begin{adjustbox}{width=\linewidth}
    \begin{tabular}{@{}llcc@{}}
    \toprule
      Dataset & Model& MAE($mm$) & MSE($mm$)  \\
    \midrule
    \multirow{2}{*}{\textit{LVQuan19}} &  ACDC      &  0.369(0.045) & 0.240(0.063) \\
    & Synthetic &  \textbf{0.332(0.038)} & \textbf{0.194(0.047)} \\
    \midrule
     \multirow{2}{*}{\textit{MS-CMRSeg}}  & ACDC      & 0.308(0.024) & 0.164(0.029) \\
    & Synthetic & \textbf{0.285(0.022)} & \textbf{0.143(0.021)} \\
    \midrule
    \multirow{2}{*}{\textit{Gen-Data}} &   ACDC      & 1.945(2.235) & 15.969(22.349) \\
    & Synthetic & \textbf{0.438(0.217)} & \textbf{0.445(0.342)} \\
    \bottomrule
    \end{tabular}
    \end{adjustbox}
\end{minipage}
\hfill
\begin{minipage}{0.49\linewidth}
    \centering
    \renewcommand{\tabcolsep}{2pt}
     \caption{
     Comparison of our proposed end-to-end model (E2ET-solver) and two baseline models (U-Net-$4$ and E2ET-gt) on three cardiac datasets. 
     }
    \label{tab:e2eresult}
    \begin{adjustbox}{width=\linewidth}
    \begin{tabular}{@{}lccc@{}}
    \toprule
        Model & \textit{ACDC} & \textit{LVQuan19} & \textit{MS-CMRSeg}\\
        \midrule
        U-Net-$4$ &  0.124(0.060) & 0.241(0.148) & 0.151(0.077) \\
        E2ET-gt & 0.071(0.046) & 0.143(0.103) & 0.086(0.053) \\ 
        E2ET-solver & \textbf{0.067(0.043)} & \textbf{0.121(0.063)} & \textbf{0.085(0.053)}\\
    \bottomrule
    \end{tabular}
    \end{adjustbox}
\end{minipage}\hfill
\end{table}

\begin{figure}
    \centering
    \includegraphics[width=0.85\textwidth]{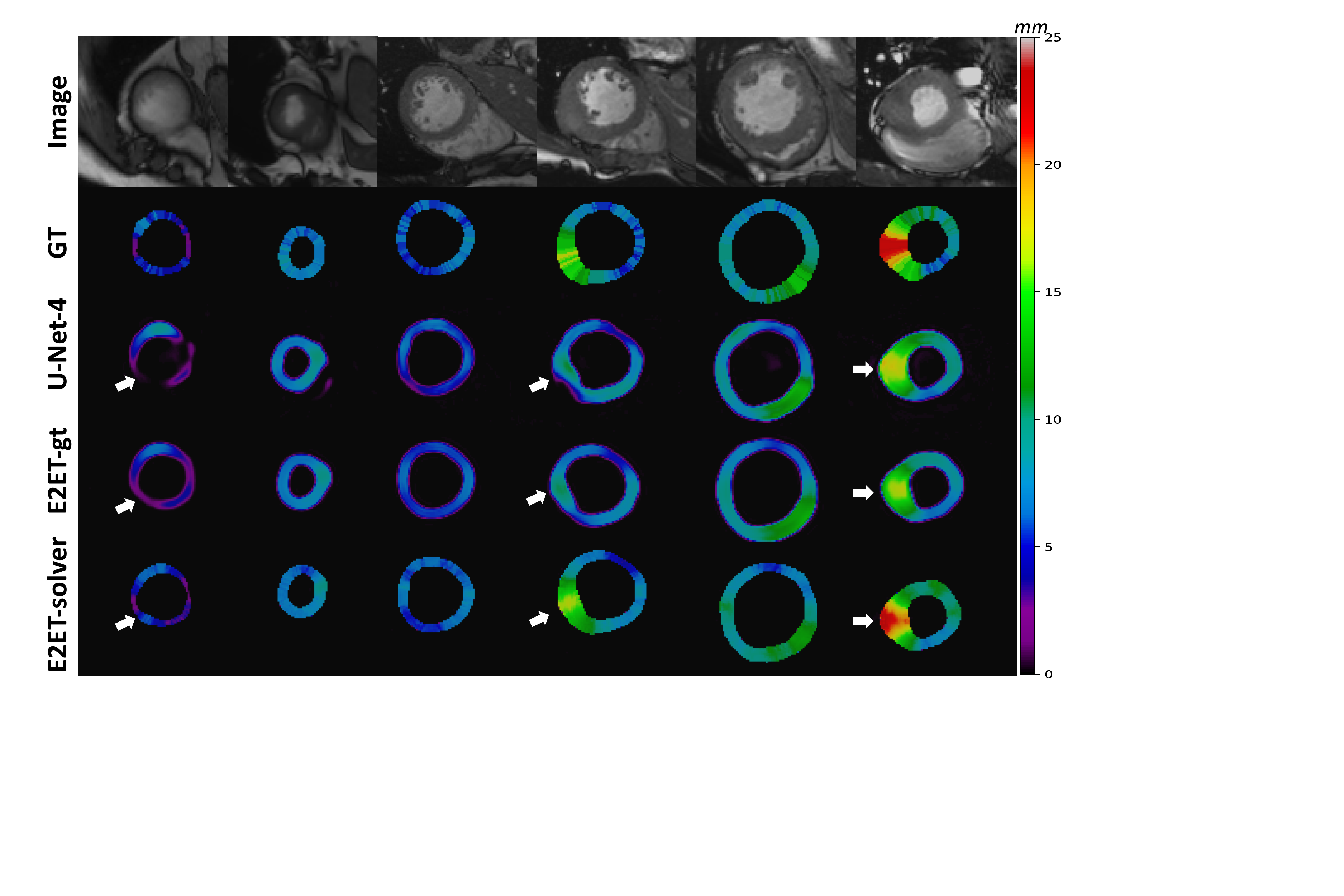}
    \caption{Examples of direct thickness prediction from images using proposed end-to-end method (E2ET-solver) and other methods (E2ET-gt and U-Net-$4$). The images are from ACDC data. The white arrows indicate the area that the E2ET-solver performs better than other models.}
    \label{fig:e2eresult}
\end{figure}

The fast thickness solver is more efficient than the mathematical model. As shown in Fig.~\ref{fig:gen}, for disconnected shapes like the two segments, the mathematical model requires the manual definition of the inner and outer boundaries while our fast thickness solver can directly calculate thickness without the manual processing.
Our solver only needs $0.35$ seconds on average for thickness prediction, around 100 times faster than the mathematical model. 

\paragraph{\textbf{Thickness Estimation on Raw Image}}\label{sec:e2e}
For comparison, we considered three end-to-end models:
(A) The baseline model U-Net-$4$ trained with loss function Eq.~\eqref{eq:ltloss}. 
(B) The proposed end-to-end model trained with loss function Eq.~\eqref{eq:newloss} but replacing $G(\hat{s})$ with ground truth thickness $y$ (E2ET-gt). 
(C) The proposed end-to-end model trained with loss function Eq.~\eqref{eq:newloss} (E2ET-solver). 
Note that E2ET-solver utilizes a fast thickness solver $G$, which is pre-trained on the synthetic dataset (see section~\ref{sec:realsolver}). 
In addition, we set $\beta=15$ in Eq.~\eqref{eq:newloss}. 
We take into account the segmentation errors so that MAE is computed on the whole image region. 
All models are trained on ACDC training set and evaluated on ACDC test set, \textit{LVQuan19} and \textit{MS-CMRSeg} dataset.

Table~\ref{tab:e2eresult} summarizes the results.
Our proposed E2ET-solver achieves the smallest MAE on all three datasets.
Fig.~\ref{fig:e2eresult} illustrates examples of thickness prediction from raw images utilizing different algorithms.
U-Net-$4$ shows large errors, which indicates the difficulties of estimating thickness from a raw image without utilizing shape information. 
E2ET-gt shows purple artifacts at the boundary regions especially for extreme challenging cases (e.g., the thin and thick regions), which may be due to the shape inconsistency as discussed in section~\ref{sec:solver_image}.
On the other hand, by disentangling the segmentation task and enforcing shape consistency, E2ET-solver shows no such artifacts and the thickness prediction is very similar to the ground truth even in the difficult cases. 

\begin{figure*}[!t]
    \centering
    \includegraphics[width=0.92\textwidth]{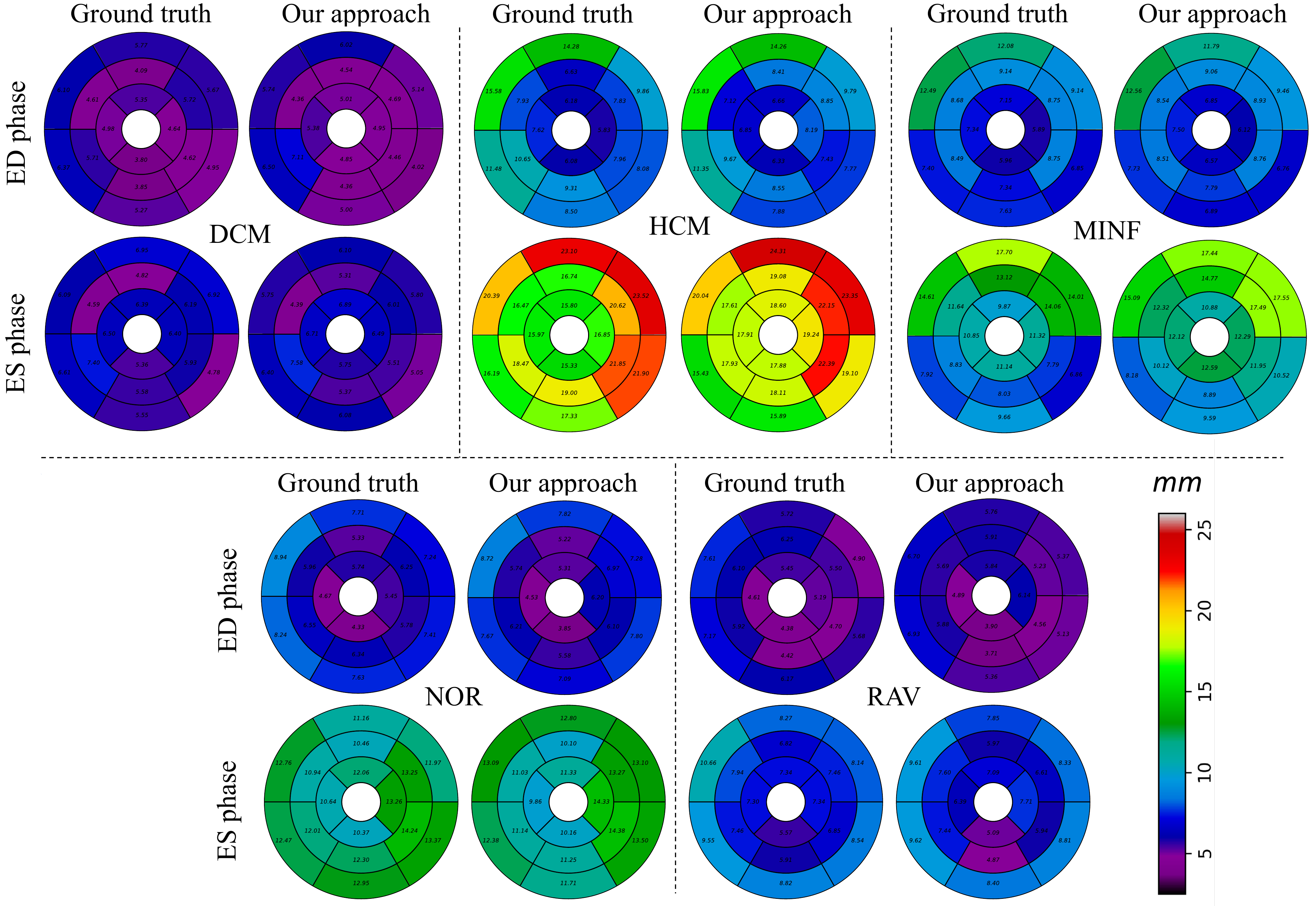}
    \caption{Examples of 17-segment regional thickness of ground truth and E2ET-solver at ED and ES phases. %
    Zoom in to see the thickness value in the middle of each segment.
    Five categories of ACDC data are displayed. 
    }
    \label{fig:path}
\end{figure*}

Qualitatively, we use the polar plot of the 17-segment model \cite{american2002standardized} to visualize different pathologies thickness patterns (can be divided into more segments for disease diagnosis) in Fig.~\ref{fig:path}.
For example, the DCM patient exhibits thin myocardium as overall small thickness values (more purple color than normal people) while the HCM patient exhibits thick myocardium as overall large thickness values (more red color than normal people).
Several thin regions can be found in the MINF patient. 
These observations are clinical symptoms of the corresponding cardiac diseases and often used as diagnostic criteria. 
Quantitatively, the E2ET-solver achieves MAE of $0.64\pm0.10~mm$, $0.73\pm0.14~mm$ and $1.08\pm0.05~mm$ in the basal, mid-cavity and apical region, respectively.
Thickness measurement near the apical region is more challenging due to the small region and the out of plane motion.
The results demonstrate the feasibility of these automated measurements for cardiac disease diagnosis. 

\section{Conclusions}
In this work, we propose an efficient and accurate thickness solver to replace the conventional mathematical model as well as a novel end-to-end model that directly estimates thickness from raw images.
Our methods demonstrate promising performance on several benchmarks.
The fast solver shows its efficiency and good generalizability on unseen shapes. 
The end-to-end model demonstrates its superior performance compared to other baselines.
We also analyze the thickness patterns of different pathologies to show the clinical value of our model in cardiac disease diagnosis.

\bibliographystyle{splncs04}
\bibliography{mybibliography}
\end{document}